\documentclass[english,conference]{IEEEtran}
\usepackage[T1]{fontenc}
\usepackage[latin9]{inputenc}
\usepackage{amsmath}
\usepackage{amsthm}
\usepackage{amssymb}
\usepackage{graphicx}

\makeatletter
\theoremstyle{plain}
\newtheorem{thm}{\protect\theoremname}
\theoremstyle{remark}
\newtheorem{rem}[thm]{\protect\remarkname}

\renewcommand\[{\begin{equation}}
\renewcommand\]{\end{equation}}

\makeatother

\usepackage{babel}
\providecommand{\remarkname}{Remark}
\providecommand{\theoremname}{Theorem}

\begin{document}
\author{
\IEEEauthorblockN{Jinwen Shi, Cong Ling} 
\IEEEauthorblockA{Imperial College London \\\{jinwen.shi12, c.ling\}@imperial.ac.uk} 
\and  \IEEEauthorblockN{Osvaldo Simeone} 
\IEEEauthorblockA{King's College London \\ osvaldo.simeone@kcl.ac.uk} 
\and   \IEEEauthorblockN{Jörg Kliewer}
\IEEEauthorblockA{New Jersey Institute of Technology\\ jkliewer@njit.edu} 
}

\title{Coded Computation Against Distributed Straggling Channel Decoders
in the Cloud for Gaussian Uplink Channels }
\maketitle
\begin{abstract}
The uplink of a Cloud Radio Access Network (C-RAN) architecture is
studied, where decoding at the cloud takes place at distributed decoding
processors. To mitigate the impact of straggling decoders in the cloud,
the cloud re-encodes the received frames via a linear code before
distributing them to the decoding processors. Focusing on Gaussian
channels, and assuming the use of lattice codes at the users, in this
paper the maximum user rate is derived such that all the servers can
reliably recover the linear combinations of the messages corresponding
to the employed linear code at the cloud. Furthermore, two analytical
upper bounds on the frame error rate (FER) as a function of the decoding
latency are developed, in order to quantify the performance of the
cloud's linear code in terms of the tradeoff between FER and decoding
latency at the cloud.
\end{abstract}

\let\thefootnote\relax

\footnote{This work has been supported in part by the Engineering and Physical
Sciences Research Council (EPSRC), the European Research Council (ERC)
under the European Union Horizon 2020 research and innovation program
(grant agreements 725731), the U.S. NSF through grant CCF-1525629,
and the U.S. NSF grant CNS-1526547.}

\section{Introduction}

\begin{figure*}[t]
\begin{centering}
\includegraphics[scale=0.8]{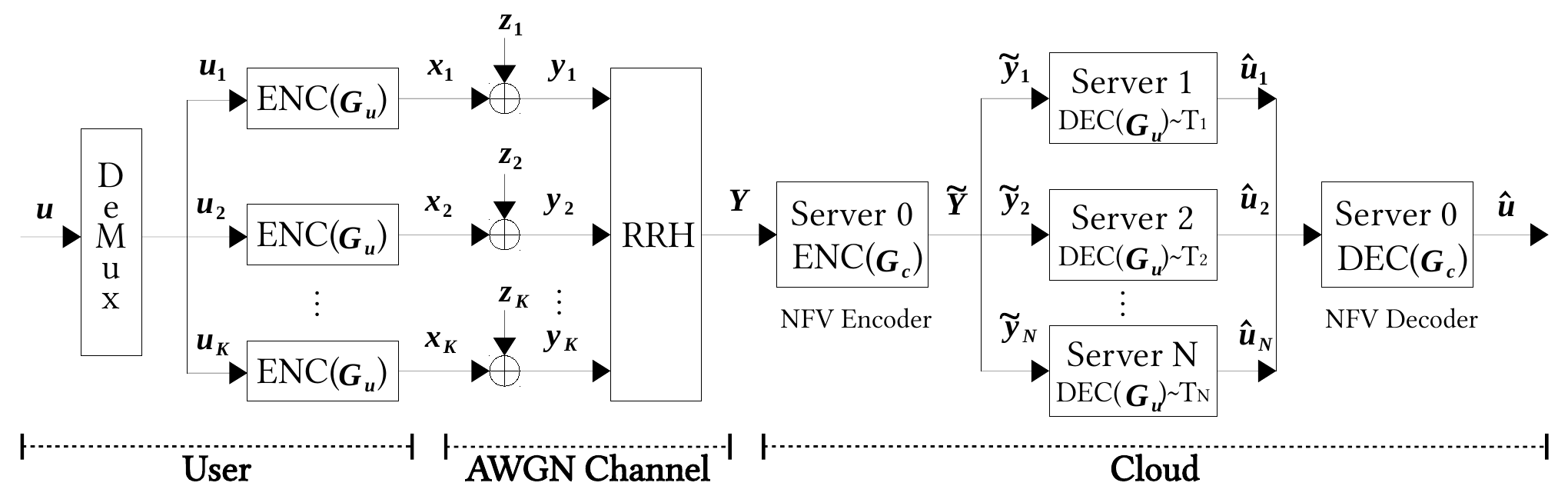}
\par\end{centering}
\caption{Distributed uplink decoding in C-RAN over an AWGN channel. \label{fig:NFV-model-for-AWGN}}
\end{figure*}

A Cloud Radio Access Network (C-RAN) architecture can leverage network
function virtualization (NFV) in order to implement baseband functionalities
on commercial off-the-shelf (COTS) hardware, such as general purpose
servers. An important challenge of this solution is to ensure a prescribed
latency performance despite the variability of the servers' runtimes
\cite{parall2018Quituna}. 

The problem of straggling processors, that is, processors lagging
behind in the execution of a certain function, has been widely studied
in the context of distributed computing \cite{DistributedComputation}.
\cite{parall2018Quituna} demonstrates the effectiveness of decomposing
tasks in parallel runnable small jobs over a distributed computing
architecture in terms of latency while avoiding overhead. 

For distributed computing, it has been recently shown in \cite{DistMachineLearning,Yang2017Computing}
that parallel processing can be improved by carrying out linear precoding
of the data prior to processing, as long as the function to be computed
is linear. The key idea is that, by employing a proper linear block
code over fractions of size $1/K$ of the original data, a function
may be completed as soon as a number of $K$ or more processors have
finalized their operation, irrespective of their identity.

The NFV-based C-RAN model considered in this paper is illustrated
by Fig. \ref{fig:NFV-model-for-AWGN}. The packets sent by a user
in the uplink are received by the remote radio head (RRH) through
an additive white Gaussian noise (AWGN) channel and forwarded to a
cloud over a RRH-to-cloud link. Decoding is carried out on a distributed
architecture consisting of COTS servers $1,\ldots,N$. 

We investigate the use of linear coding on the received packets as
a means to improve over parallel processing in order to mitigate the
impact of straggling decoders at the cloud. The idea was first studied
in \cite{Malihe2017NFV,NFV2016SimeoneKliewer} where the packets are
received by the RRH via a binary symmetric channel (BSC). In this
paper, we tackle the problem of extending the design and analysis
to Gaussian channels. 

With Gaussian channels, the model at hand is similar to the compute-and
forward (C\&F) problem \cite{ComptForward} emerging in Gaussian relay
networks. In this problem, the relays attempt to decode their received
signals into integer linear combinations of codewords, which they
then forward to the destinations. The main difference is that in the
C\&F transmitted signals are mixed by the channel, while in our model
linear combining is applied at the cloud. Accordingly, in the NFV
scenarios, the linearly combined received packets contain an accumulated
noise term (i.e., $\mathbf{\tilde{y}}_{i}=\sum_{j=1}^{K}a_{ij}\left(\mathbf{x}_{j}+\mathbf{z}_{j}\right)$),
while this is not the case in C\&F setting (i.e., $\mathbf{\tilde{y}}_{i}=\sum_{j=1}^{K}a_{ij}\mathbf{x}_{j}+\mathbf{z}_{j}$). 

The accumulated noise terms (i.e., $\sum_{j=1}^{K}a_{ij}\mathbf{z}_{j}$)
affect the functions of the servers in terms of the following two
aspects. First, noise powers are accumulated, which leads to a variation
on the decoding error probability of each individual server compared
to the C\&F problem. Second, the common terms in $\sum_{j=1}^{K}a_{ij}\mathbf{z}_{j}$
make the noise terms seen by the servers in general dependent. 

To account for the first aspect, we derive the computation rate that
guarantees correct decoding for each server in Sec. \ref{sec:Analytical-Bounds-on-FER}.
As for the second aspect, we analyze the dependency among the servers
by using the dependency graph of the linear NFV code as introduced
in \cite{Malihe2017NFV}. Then, we derive two analytical upper bounds
on the frame error rate (FER) as a function of the decoding latency.
The bounds on FER depend on the properties of both the channel coding
adopted by the user and the linear NFV code applied at the cloud.

\textit{Notation:} Let $+$, $\sum$ and $\oplus$, $\bigoplus$ denote
addition and summation over reals and finite fields, respectively.
Let $\left\Vert \mathbf{h}\right\Vert \triangleq\sqrt{\sum_{i=1}^{N}\left|h_{i}\right|^{2}}$
denote the norm of a vector $\mathbf{h}$. $\left[K\right]$ denotes
the set $\left\{ 1,2,\cdots,L\right\} $. All logarithms are of base
two. Let $\log^{+}\left(x\right)\triangleq\max\left(\log\left(x\right),0\right)$.
$\left|\mathcal{F}\right|$ denote cardinality of $\mathcal{F}$. 

\section{Problem Statement}

\subsection{System Model}

As illustrated in Fig. \ref{fig:NFV-model-for-AWGN}, we focus on
the uplink of a C-RAN system with a multi-server cloud decoder connected
to an RRH via a dedicated fronthaul link. As detailed next, the model
follows reference \cite{Malihe2017NFV}, but it considers the more
realistic AWGN channel for the user-RRH link, requiring a redesign
of the operation at the cloud.

The user encodes a file $\mathbf{u}$ of length $L$ over a finite
field $\mathbb{F}_{p}$ for uplink transmission, where $p>0$ is a
prime in $\mathbb{Z}$. Each symbol is drawn independently and uniformly
over the finite field. Before encoding, the file is divided into $K$
blocks $\mathbf{u}_{1}$, $\mathbf{u}_{2}$, $\ldots$, $\mathbf{u}_{K}$
of equal length $k\triangleq L/K$ symbols. The user's encoder, $\mathcal{E}:\mathbb{F}_{p}^{k}\rightarrow\mathbb{R}^{n}$,
then maps each length-$k$ block to a length-$n$ real valued codeword,
$\mathbf{x}_{j}=\mathcal{E}\left(\mathbf{u}_{j}\right)$. The encoder
is subject to the power constraint $\mathbb{E}[\left\Vert \mathbf{x}_{j}\right\Vert ^{2}]\leq nP.$
The transmission rate $R$ of the user is the length of its message
normalized by the number of channel uses, i.e., $R=k/n\log p.$

At the output of the user-RRH AWGN channel, the length-$n$ received
packet for the $j$-th block at the RRH is given as 
\begin{equation}
\mathbf{y}_{j}=\mathbf{x}_{j}+\mathbf{z}_{j},\label{eq:ChannelModel}
\end{equation}
where $\mathbf{z}_{j}$ is a vector of i.i.d. Gaussian random variables
with zero-mean and variance $N_{0}$. For convenience, we define the
signal-to-noise ratio (SNR) as $\textrm{SNR}\triangleq P/N_{0}$.
The $K$ packets $\left(\mathbf{y}_{1},\mathbf{y}_{2},\ldots,\mathbf{y}_{K}\right)$
are transmitted by the RRH to the cloud over a fronthaul link. Decoding
is carried out at the cloud. 

To this end, the cloud consists of $N$ available servers, namely,
Server $1,\ldots,N$, and a master server, i.e., Server $0$. Each
server can decode a packet within a random time $T_{i}=T_{1,i}+T_{2,i}$,
where times $\left\{ T_{1},\ldots,T_{N}\right\} $ are mutually independent.
Time $T_{1,i}$ accounts the unavailability of the processor, and
is independent of the workload, while $T_{2,i}$ models the execution
runtime and it grows as the size $n$ of the packet. The variable
$T_{1,i}$ follows an exponential distribution with mean $1/\mu_{1}$,
while $T_{2,i}$ is a shifted exponential with shift equal to $a\geq0$
and average equal to $a+1/\mu_{2}\times n$ so that $1/\mu_{2}$ is
the time required for an input symbol. The probability that a given
set of $l$ out of $N$ servers has finished decoding by time $t$
is given as $\textrm{Pr}\left(l,t\right)=F\left(t\right)^{l}\left(1-F\left(t\right)\right)^{N-l}$,
where $F\left(t\right)$ is the cumulative distribution function of
$T_{i}$.

In order to mitigate the effect of decoding straggling, we adapt the
NFV coding scheme in \cite{Malihe2017NFV} to the AWGN channel. NFV
coding operates as follows. The $K$ packets are first linearly encoded
by Server $0$ into $N\geq K$ coded blocks of the same length $n$,
as depicted in Fig. \ref{fig:NFV-model-for-AWGN}. The reason for
this partitioning is that each block is forwarded to a different server
in the cloud for decoding. For linear coding, consider an $\left(N,K\right)$
linear code $\mathcal{C}_{c}$ with $K\times N$ generator matrix
$\mathbf{G}_{c}\in g\left(\mathbb{\mathbb{F}}_{p'}\right)^{N\times K}$,
where $p'>0$ is a prime and $g\left(\cdot\right)$ is the natural
map from $\mathbb{F}_{p'}$ to the integers $\left\{ 0,1,2,\ldots,p'-1\right\} $.
Note that the prime $p'$ may be different from the prime $p$ used
to define the user code. Accordingly, the encoded packets are obtained
as 
\[
\tilde{\mathbf{Y}}=\mathbf{Y}\mathbf{G}_{c},
\]
where $\mathbf{Y}=\left[\mathbf{y}_{1},\mathbf{y}_{2},\ldots,\mathbf{y}_{K}\right]$
is a $n\times K$ matrix, and $\tilde{\mathbf{Y}}=\left[\mathbf{\tilde{y}}_{1},\mathbf{\tilde{y}}_{2},\ldots,\mathbf{\tilde{y}}_{N}\right]$
is a $n\times N$ matrix. From (\ref{eq:ChannelModel}), the encoded
packet $\mathbf{\tilde{y}}_{i}$ can be written as 
\begin{equation}
\mathbf{\tilde{y}}_{i}=\sum_{j=1}^{K}\mathbf{y}_{j}g_{c,ji}=\sum_{j=1}^{K}\mathbf{x}_{j}g_{c,ji}+\sum_{j=1}^{K}\mathbf{z}_{j}g_{c,ji},\label{eq:reveivedPacket}
\end{equation}
where $g_{c,ji}$ is the $\left(j,i\right)$ entry of matrix $\mathbf{G}_{c}$. 

Each server $i\in\left[N\right]$ aims at decoding a linear combination
of the messages 
\begin{equation}
\mathbf{\tilde{u}}_{i}=\bigoplus_{j=1}^{K}\tilde{g}_{c,ji}\mathbf{u}_{j},\label{eq:linear-combination-messages}
\end{equation}
where $\tilde{g}_{c,ji}=g^{-1}\left(\left[g_{c,ji}\right]\textrm{ mod }p\right)$
are coefficients taking values in $\mathbb{F}_{p}$. To this end,
Server $i$ is equipped with a decoder, $\mathcal{D}_{i}:\mathbb{R}^{n}\rightarrow\mathbb{F}_{p}^{k}$,
that maps the observed output $\mathbf{\tilde{y}}_{i}$ to an estimate
$\mathbf{\hat{u}}_{i}=\mathcal{D}_{i}\left(\mathbf{\tilde{y}}_{i}\right)$
of the equation $\mathbf{\tilde{u}}_{i}$. 

Let $d_{\textrm{min}}$ be the minimum distance of the NFV code $\mathcal{C}_{c}$.
Server $0$ is able to decode the message $\mathbf{u}$, or equivalently
the $K$ packets $\mathbf{u}_{j}$ for $j\in\left[K\right]$, as soon
as $N-d_{\textrm{min}}+1$ servers have decoded successfully. The
output $\mathbf{\hat{u}}_{i}\left(t\right)$ at the $i$th Server
at time $t$ is $\mathbf{\hat{u}}_{i}\left(t\right)=\mathbf{\hat{u}}_{i}$,
if $T_{i}\leq t$; and $\mathbf{\hat{u}}_{i}\left(t\right)=\emptyset$,
otherwise. The output $\mathbf{\hat{u}}\left(t\right)$ of the decoder
at Server $0$ at time $t$ is a function of $\mathbf{\hat{u}}_{i}\left(t\right)$
for $i\in\left[N\right]$. The frame error rate (FER) at time $t$
is defined as 
\begin{equation}
P_{e}^{\textrm{FER}}\left(t\right)=\textrm{Pr}\left(\mathbf{\hat{u}}\left(t\right)\neq\mathbf{u}\right).\label{eq:FER-def}
\end{equation}

\section{Analytical Bounds on the FER \label{sec:Analytical-Bounds-on-FER}}

In this section we study the trade-off between the decoding latency
and the decoding error probability, by deriving an upper bound on
the FER $P_{e}^{\textrm{FER}}\left(t\right)$ in (\ref{eq:FER-def}). 

Each Server $i$ with $i\in\left[N\right]$ outputs the correct equation
$\mathbf{\tilde{u}}_{i}$ by time $t$ if: \textit{(i)} the server
completes decoding at time $t$, and \textit{(ii)} the decoder can
correctly decode despite the noise caused by the AWGN channel. We
define the indicator variables $C_{i}\left(t\right)=\mathbf{1}\left\{ T_{i}\leq t\right\} $
and $D_{i}\left(t\right)=\mathbf{1}\left\{ \mathbf{\hat{u}}_{i}=\mathbf{\tilde{u}}_{i}\right\} $,
which equal $1$ if the above two events occur, respectively, and
zero otherwise. Recalling that an error occurs at time $t$ if the
number of servers that have successfully decoded by time $t$ is smaller
than $N-d_{\textrm{min}}+1$. With these definitions, the FER is given
by 
\[
P_{e}^{\textrm{FER}}\left(t\right)=\Pr\left(\sum_{i=1}^{N}C_{i}\left(t\right)D_{i}\left(t\right)\leq N-d_{\textrm{min}}\right).
\]

The variables $C_{i}\left(t\right)$ are independent Bernoulli random
variables across the servers $i\in\left[N\right]$, due to the independence
among the decoding times $\left\{ T_{i}\right\} _{1}^{N}$. However,
the variables $D_{i}\left(t\right)$ are dependent Bernoulli random
variables, since there may exist common terms among the noise terms
$\sum_{j=1}^{K}\mathbf{z}_{j}g_{c,ji}$ in (\ref{eq:reveivedPacket})
at the decoders. The dependency of variables $D_{i}\left(t\right)$
is accounted for when deriving an the upper bound on the FER shown
in Sec. \ref{subsec:Upper-Bounds-on-FER}.

In order to compute an upper bound on the FER, we first evaluate the
computation rate, which gives the maximum rate for each Server $i$
to decode the desired equation $\mathbf{\tilde{u}}_{i}$ with average
probability of error approaching zero. Based on this auxiliary result,
we then employ the error exponent given in \cite[Theorems 8-11]{FDIC2013Zamir}
to characterize the upper bounds on the decoding error probability
of each Server $i$ under a given coefficient vector $\mathbf{g}_{c,i}$
and a given SNR. Finally, we give two upper bounds on the FER by taking
account the combined impact from the dependence of $D_{i}\left(t\right)$
and the accumulated noise.

\subsection{Computation Rate \label{sec:Computation-Rate}}

In order to allow servers to decode the desired equations in a manner
similar to C\&F, we assume that the user adopts a nested lattice code.
In this subsection, we derive conditions on the NFV code that enable
the servers to decode the desired equations.

To proceed, the following definitions are useful. An $n$-dimensional
lattice is a discrete subgroup of $\mathbb{R}^{n}$ which can be described
by 
\[
\Lambda=\{\mathbf{\lambda}=\mathbf{B}\mathbf{z}:\textrm{ }\mathbf{z}\in\mathbb{Z}^{n}\},
\]
where $\mathbf{B}$ is the full rank generator matrix. The Voronoi
region $\mathcal{V}$ of a lattice $\Lambda$ is 
\[
\mathcal{V}\triangleq\left\{ \mathbf{z}:Q_{\Lambda}\left(\mathbf{z}\right)=\mathbf{0}\right\} ,
\]
where $Q_{\Lambda}\left(\mathbf{z}\right)\triangleq\arg\min_{\mathbf{\lambda}\in\Lambda}\left\Vert \mathbf{z}-\mathbf{\lambda}\right\Vert $.
Let $\textrm{Vol}\left(\mathcal{V}\right)$ denote the volume of $\mathcal{V}$
and $\textrm{Vol}\left(\mathcal{V}\right)=\left|\det\left(\mathbf{B}\right)\right|$.
The second moment of a lattice $\Lambda$ is defined as 
\[
\sigma_{\Lambda}^{2}\triangleq\frac{1}{n\textrm{Vol}\left(\mathcal{V}\right)}\int_{\mathcal{V}}\left\Vert \mathbf{\mathbf{z}}\right\Vert ^{2}d\mathbf{\mathbf{z}},
\]
and the normalized second moment (NSM) is defined as 
\begin{equation}
G\left(\Lambda\right)\triangleq\frac{\sigma_{\Lambda}^{2}}{\left(\textrm{Vol}\left(\mathcal{V}\right)\right)^{2/n}}.\label{eq:Def-NSM}
\end{equation}
A lattice $\Lambda$ is said to be nested in a lattice $\Lambda_{f}$
if $\Lambda\subseteq\Lambda_{f}$. Refer $\Lambda_{f}$ as the fine
lattice and $\Lambda$ as the coarse lattice.

The following theorem provides a condition on the transmission rate
$R$ that guarantees reliable decoding of given equations at the servers.
\begin{thm}
\label{thm:ComputationRate} For a given NFV code matrix $\mathbf{G}_{c}$
and $n$ large enough, there exists a nested lattice code $\Lambda\subseteq\Lambda_{f}$
with rate $R$, such that for all coefficient vectors $\mathbf{g}_{c,1}$,
$\mathbf{g}_{c,2}$,$\ldots$, $\mathbf{g}_{c,N}\in g\left(\mathbb{\mathbb{F}}_{p'}\right)^{K}$,
any Server $i\in\left[N\right]$ can recover the linear combination
of messages $\mathbf{\tilde{u}}_{i}$ given in (\ref{eq:linear-combination-messages})
with average probability of error $\epsilon$ as long as the inequality
\[
R<\min_{i:g_{c,ji}\neq0}\frac{1}{2}\log^{+}\left(\frac{P}{\left\Vert \mathbf{g}_{c,i}\right\Vert ^{2}N_{0}\left(\alpha_{i}^{2}+\textrm{SNR}\left(\alpha_{i}-1\right)^{2}\right)}\right)
\]
holds for some choice of parameters $\alpha_{1},\ldots,\alpha_{N}\in\mathbb{R}$. 
\end{thm}
\begin{IEEEproof}
See Appendix A.
\end{IEEEproof}
Based on Theorem \ref{thm:ComputationRate}, we define the computation
rate for each Server $i$ as

\begin{equation}
\mathcal{R}^{*}\left(\mathbf{g}_{c,i}\right)=\max_{\alpha_{i}\in\mathbb{R}}\frac{1}{2}\log^{+}\left(\frac{P}{\left\Vert \mathbf{g}_{c,i}\right\Vert ^{2}N_{0}\left(\alpha_{i}^{2}+\textrm{SNR}\left(\alpha_{i}-1\right)^{2}\right)}\right).\label{eq:computation-rate}
\end{equation}
By Theorem \ref{thm:ComputationRate}, this is the rate that guarantees
correct decoding at Server $i$.

\begin{thm}
\label{thm:optimalCompuRate}The computation rate (\ref{eq:computation-rate})
is uniquely maximized by choosing $\alpha_{i}$ to be the minimum
mean square error (MMSE) coefficient $\alpha_{MMSE}=\frac{\textrm{SNR}}{1+\textrm{SNR}}$
which results in a computation rate of 
\[
\mathcal{R}^{*}\left(\mathbf{g}_{c,i}\right)=\frac{1}{2}\log^{+}\left(\frac{1+\textrm{SNR}}{\left\Vert \mathbf{g}_{c,i}\right\Vert ^{2}}\right).
\]
\end{thm}
\begin{IEEEproof}
See Appendix B.
\end{IEEEproof}
\begin{rem}
The computation rate from Theorem \ref{thm:optimalCompuRate} is zero
if the coefficient vector $\mathbf{g}_{c,i}$ satisfies $\left\Vert \mathbf{g}_{c,i}\right\Vert ^{2}\geq1+\textrm{SNR}$.
\end{rem}

\subsection{Upper Bounds on the FER \label{subsec:Upper-Bounds-on-FER}}

In order to analyze the FER, we need to first evaluate the decoding
error probability for each Server $i$, for $i\in\left[N\right]$,
as a function of the vector $\mathbf{g}_{c,i}$ defined by the NFV
code. 

To this end, define the gap to the computation rate as
\[
\Delta=\frac{1}{2}\log^{+}\left(\frac{1+\textrm{SNR}}{\left\Vert \mathbf{g}_{c,i}\right\Vert ^{2}}\right)-R,
\]
and let $\mu\triangleq2^{2\Delta}$. Assuming maximum likelihood (ML)
decoding, an upper bound on the decoding error probability is given
by $P_{e}^{\textrm{ML}}\left(\mathbf{g}_{c,i}\right)$ \cite[Theorems 8-11]{FDIC2013Zamir},
where

\begin{multline}
\begin{split}P_{e}^{\textrm{ML}}\left(\mathbf{g}_{c,i}\right)\cong & \begin{cases}
e^{-nE_{r}\left(\mu\right)}\frac{1}{\sqrt{2\pi n}}, & \mu>2\\
e^{-nE_{r}\left(\mu\right)}\frac{1}{\sqrt{8\pi n}}, & \mu=2\\
\frac{e^{-nE_{r}\left(\mu\right)}\left(n\pi\right)^{-\frac{\mu}{2}}}{\left(2-\mu\right)\left(\mu-1\right)}, & 2>\mu>1,
\end{cases}\end{split}
\label{eq:ErrProbBound_ML}
\end{multline}
where $a\cong b$ indicates that $\frac{a}{b}\rightarrow1$, and $E_{r}\left(\cdot\right)$
is the Poltyrev random coding exponent defined as \cite{poltyrev} 

\begin{multline}
\begin{split}E_{r}\left(\mu\right)= & \begin{cases}
\frac{1}{2}\left[\ln\left(\mu\right)+\ln\left(e/4\right)\right], & \mu\geq2\\
\frac{1}{2}\left[\mu-1-\ln\left(\mu\right)\right], & 2\geq\mu\geq1\\
0, & \mu\leq1.
\end{cases}\end{split}
\label{eq:poltyrev-exponent}
\end{multline}

Based on the bound (\ref{eq:ErrProbBound_ML}), we now provide an
upper bound on the FER by leveraging the approach introduced in \cite{Malihe2017NFV}.
Accordingly, we use the notion of the dependence graph and its chromatic
number for the NFV code to characterize the dependence of the correct
decoding indications $D_{i}$. 

The dependence graph $\mathcal{G}\left(\mathbf{G}_{c}\right)=\left(\Upsilon,\mathcal{\varOmega}\right)$
comprises a set $\Upsilon$ of $N$ vertices and a set $\mathcal{\varOmega}\subseteq\Upsilon\times\Upsilon$
of edges, where the edge $\left(i,j\right)\in\mathcal{\varOmega}$
is included if both the $i$th and $j$th columns of $\mathbf{G}_{c}$
have at least a non-zero term in the same row. Each vertex of $\mathcal{G}\left(\mathbf{G}_{c}\right)$
represents a decoding server, and an edge indicates that the noise
terms in (\ref{eq:reveivedPacket}) for the two servers are correlated.
The chromatic number $\mathcal{X}\left(\mathbf{G}_{c}\right)$ of
$\mathcal{G}\left(\mathbf{G}_{c}\right)$ is the smallest number of
colors needed to color the vertices of $\mathcal{G}\left(\mathbf{G}_{c}\right)$,
such that no two adjacent vertices share the same color. We then give
a large deviation bound (LDB) on the FER. 
\begin{thm}
\cite[Theorem 1]{Malihe2017NFV}\label{thm:upBoundFER} Let $P_{e}^{\min}=\min_{i}\left\{ P_{e}^{\textrm{ML}}\left(\mathbf{g}_{c,i}\right)\right\} _{i=1}^{N}$,
according to (\ref{eq:ErrProbBound_ML}). Then, for all $t\geq n\left(a-\frac{1}{\mu}\ln\left(\frac{d_{\textrm{min}}-\sum_{i=1}^{N}P_{e}^{\textrm{ML}}\left(\mathbf{g}_{c,i}\right)}{N-\sum_{i=1}^{N}P_{e}^{\textrm{ML}}\left(\mathbf{g}_{c,i}\right)}\right)\right),$
the FER is upper bounded as 
\[
\begin{split} & P_{e}^{\textrm{FER}}\left(t\right)\leq\exp\left(-\frac{S\left(t\right)}{b^{2}\left(t\right)\mathcal{X}\left(\mathbf{G}_{c}\right)}\right.\\
 & \left.\hspace{-5bp}\cdot\varphi\left(\frac{4b\left(t\right)\left(NF\left(t\right)-F\left(t\right)\sum_{i=1}^{N}P_{e}^{\textrm{ML}}\left(\mathbf{g}_{c,i}\right)-N+d_{\min}\right)}{5S\left(t\right)}\right)\right),
\end{split}
\]
where $b\left(t\right)\triangleq F\left(t\right)\left(1-P_{e}^{\min}\right)$,
$S\left(t\right)\triangleq\sum_{i=1}^{N}F\left(t\right)\left(1-P_{e}^{\textrm{ML}}\left(\mathbf{g}_{c,i}\right)\right)\left(1-F\left(t\right)\left(1-P_{e}^{\textrm{ML}}\left(\mathbf{g}_{c,i}\right)\right)\right)$,
and $\varphi\left(x\right)\triangleq\left(1+x\right)\ln\left(1+x\right)-x$. 
\end{thm}
This upper bound captures the dependency of the FER caused by the
NFV code, and also the error probability $P_{e}^{\textrm{ML}}\left(\mathbf{g}_{c,i}\right)$
depending on both the channel code and the NFV code. The following
gives a union bound (UB) that is tighter and valid for all times $t$.
\begin{thm}
\cite[Theorem 2]{Malihe2017NFV} For any subset $\mathcal{A}\subseteq\left[N\right]$,
define $P_{e}^{\min\left(\mathcal{A}\right)}\triangleq\min_{i}\left\{ P_{e}^{\textrm{ML}}\left(\mathbf{g}_{c,i}\right)\right\} _{i\in\mathcal{A}}$
and $P_{e}^{\mathcal{A}}\triangleq\sum_{i\in\mathcal{A}}P_{e}^{\textrm{ML}}\left(\mathbf{g}_{c,i}\right)$,
and let $\mathbf{G}_{\mathcal{A}}$ be the $K\times\left|\mathcal{A}\right|$,
submatrix of $\mathbf{G}_{c}$, with column indices in the subset
$\mathcal{A}$. Then, the FER is upper bounded by 
\begin{multline*}
\begin{split} & P_{e}^{\textrm{FER}}\left(t\right)\leq1-\hspace{-10bp}\sum_{l=N-d_{\min}+1}^{N}\hspace{-10bp}\textrm{Pr}\left(l,t\right)\hspace{-10bp}\sum_{\mathcal{A}\subseteq\left[N\right]:\left|\mathcal{A}\right|=l}\left(1-\right.\\
 & \left.\exp\left(-\frac{S_{\mathcal{A}}}{b_{\mathcal{A}}^{2}\mathcal{X}\left(\mathbf{G}_{\mathcal{A}}\right)}\varphi\left(\frac{4b_{\mathcal{A}}\left(l-N+d_{\min}-P_{e}^{\mathcal{A}}\right)}{5S_{\mathcal{A}}}\right)\right)\right),
\end{split}
\end{multline*}
where $S_{\mathcal{A}}\left(t\right)\triangleq\sum_{i\in\mathcal{A}}P_{e}^{\textrm{ML}}\left(\mathbf{g}_{c,i}\right)\left(1-P_{e}^{\textrm{ML}}\left(\mathbf{g}_{c,i}\right)\right)$
and $b_{\mathcal{A}}\triangleq1-P_{e}^{\min\left(\mathcal{A}\right)}$.
\end{thm}

\section{Numerical Results}

In this section, we provide some numerical results to obtain insights
into the performance of NFV codes based on the FER bounds presented
in the previous section, in terms of the trade-offs between decoding
latency and FER. We employ a frame length of $L=504$ and $N=8$ servers.
The user code is selected to be binary (i.e., $p=2$) with rate $R=0.5$.
We set $\mu_{1}=50$, $\mu_{2}=10$, and $a=1$. Unless stated, otherwise,
we have $p'=p=2$. Furthermore, we leave the performance comparison
with simulated results based on specific user lattice codes to future
work (See \cite{Malihe2017NFV} for the case of binary symmetric channels).

\begin{figure}
\begin{centering}
\includegraphics[width=1\columnwidth]{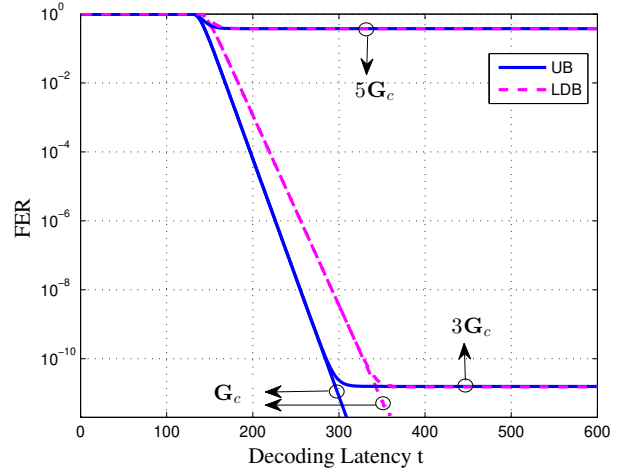}
\par\end{centering}
\caption{Comparison of LDB and UB based on ML decoding for parallel processing,
whose generator matrices are set to be $\mathbf{G}_{c}\triangleq\mathbf{I}^{N\times N}$,
$3\mathbf{G}_{c}$, and $5\mathbf{G}_{c}$. ($L=504$, $N=8$, $\mu_{1}=50$,
$\mu_{2}=10$, $a=1$, $p=2$, $p'=\left\{ 2,5,7\right\} $, $\textrm{SNR}=18\textrm{ dB}$)
\label{fig:Sim_LDB_UB_NFV_RC_ML}}
\end{figure}

We compare the performance of the following solutions: \textit{(i)
Single-server (SS) decoding}, where there is a single server $N=1$
at the cloud that decodes the entire frame $(K=1)$, so that we have
$n=1008$ and $\mathcal{X}\left(\mathbf{G}_{c}\right)=d_{\textrm{min}}=1$;
\textit{(ii) Repetition coding (RPT)}, where the entire frame is duplicated
at all servers, so that we have $n=1008$ and $\mathcal{X}\left(\mathbf{G}_{c}\right)=d_{\textrm{min}}=8$;
\textit{(iii) Parallel processing (PRL)}, where the frame is divided
into $K=N$ disjoint parts processed by different servers in parallel,
and hence we have $n=126$ and $\mathcal{X}\left(\mathbf{G}_{c}\right)=d_{\textrm{min}}=1$;
\textit{(iv) Single parity check code (SPC)}, with $K=7$, where one
servers decodes a sum of all other $K$ received packets, and hence
we have $n=144$ and $\mathcal{X}\left(\mathbf{G}_{c}\right)=d_{\textrm{min}}=2$;
and \textit{(v) an NFV code} $\mathcal{C}_{c}$ with generator matrix
$\mathbf{G}_{c}$ defined in \cite[Eq. (8)]{Malihe2017NFV} which
is characterized by $K=4$, $n=252$ and $\mathcal{X}\left(\mathbf{G}_{c}\right)=d_{\textrm{min}}=3$. 

In order to elaborate on the optimal computation rate in Theorem \ref{thm:optimalCompuRate},
Figure \ref{fig:Sim_LDB_UB_NFV_RC_ML} shows the LDB and UB for three
parallel coding schemes with generator matrices $\mathbf{G}_{c}=\mathbf{I}^{N\times N}$,
$3\mathbf{G}_{c}$, and $5\mathbf{G}_{c}$. Note that all these parallel
codes have the same minimum Hamming distance $d_{\min}=1$ and the
same chromatic number $\mathcal{X}\left(\mathbf{G}_{c}\right)=1$,
since the positions of all the non-zeros elements are the same. However,
they take entries from different field sizes, e.g., $p'=2,5,7$. Figure
\ref{fig:Sim_LDB_UB_NFV_RC_ML} confirms the main result in Theorem
\ref{thm:optimalCompuRate} that, under the same $\textrm{SNR}$,
the NFV codes with larger norms on the column vectors of the generator
matrix entails a larger equivalent noise for the server to decode
the message equations, causing a larger error floor, and accordingly,
a worse trade-off between latency and FER. Larger fields may offer
opportunities for the design of more efficient codes, which we leave
as an open problem.

\begin{figure}
\begin{centering}
\includegraphics[width=1\columnwidth]{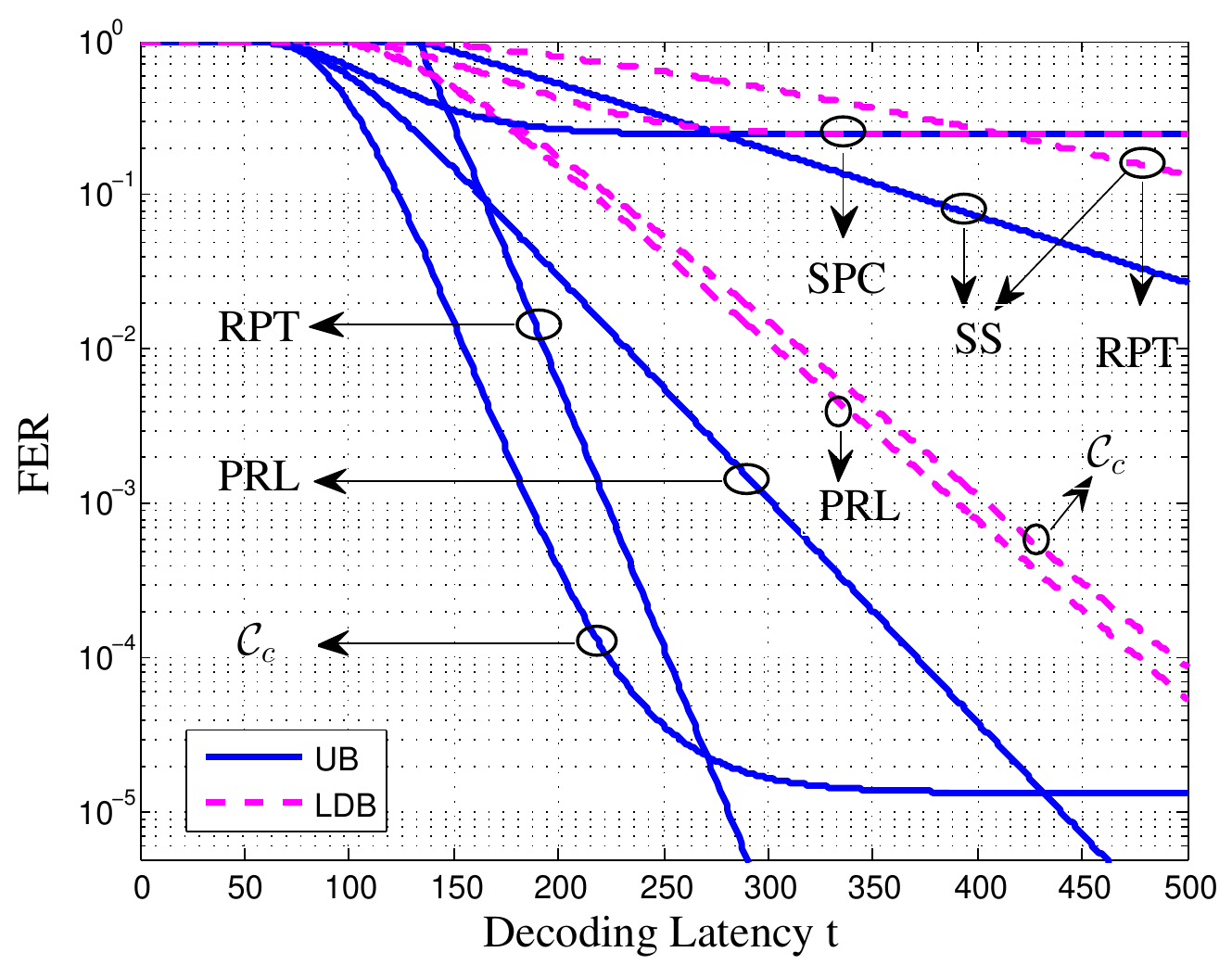}
\par\end{centering}
\caption{LDB and UB based on ML decoding for single-server decoding (SS), repetition
coding (RPT), parallel processing (PRL), single parity-check code
(SPC) and the NFV code $\mathcal{C}_{c}$ defined by $\mathbf{G}_{c}$
given in \cite[Eq. (8)]{Malihe2017NFV}. ($L=504$, $N=8$, $\mu_{1}=1/30$,
$\mu_{2}=10$, $a=0.1$, $p=2$, $p'=2$, $\textrm{SNR}=7\textrm{ dB}$)\label{fig:Sim-Multi-schemes-Unavailble-SNR5}}

\end{figure}

To compare different NFV coding schemes, Figure \ref{fig:Sim-Multi-schemes-Unavailble-SNR5}
is obtained with parameters $\mu_{1}=1/30$, $\mu_{2}=10$, and $a=0.1$,
in which we consider the case where latency may be dominated by effects
that are independent of $n$, i.e., $\mu_{1}=1/30$. Figure \ref{fig:Sim-Multi-schemes-Unavailble-SNR5}
shows both LDB and UB for all the five schemes under $\textrm{SNR}=7\textrm{ dB}$.
As first observation, Figure \ref{fig:Sim-Multi-schemes-Unavailble-SNR5}
confirms that UB is tighter than the LDB, and we note that leveraging
multiple servers for decoding yields a better trade-off between latency
and FER. 

Figure \ref{fig:Sim-Multi-schemes-Unavailble-SNR5} shows that, according
to the derived upper bounds, the NFV code $\mathcal{C}_{c}$ provides
the smallest FER for a sufficiently small latency level, improving
over all schemes including parallel processing. The latter scheme
is in fact very sensitive to the unavailability of the servers, requiring
all servers to complete decoding, and hence it needs a longer latency
in order to achieve a low FER. As for the SPC scheme, although it
has an extra parity-check server as compared to parallel processing,
its performance is limited by the large equivalent noise determined
by its coding matrix. We emphasize that these conclusions are drawn
based solely on the derived upper bound, but simulation results for
practical codes are expected to show a similar behavior (see \cite{DistMachineLearning}).

\section{Conclusion}

In this work, we have extended the idea of coding to improve the robustness
of uplink channel decoding in the cloud over AWGN channels. Explicit
calculations on the computation rate are provided to quantify the
impact on the accumulated noise terms caused by linear coding over
the received packets. Taking account the dependency among servers
and the equivalent noise for each server, we have derived upper bounds
on the FER depending on both the channel coding and the NFV coding,
and evaluate the trade-offs between FER and decoding latency under
various coding schemes. As future work, we mention here the optimized
design of NFV codes as a function of the field size.

\bibliographystyle{IEEEtran}
\bibliography{MyreffGeneral}

\section*{Appendix A}

The user's encoder $\mathcal{E}$ maps its finite field message vector
$\mathbf{u}_{j}$ to a lattice point $\mathbf{t}_{j}\in\Lambda_{f}\cap\mathcal{V}$,
using the function $\phi$ from \cite[Lemma 5]{ComptForward}, i.e.,
$\mathbf{t}_{j}=\phi\left(\mathbf{u}_{j}\right)$. In order to recover
$\mathbf{\tilde{u}}_{i}$, each Server $i$ needs to decode the lattice
equation 
\[
\mathbf{v}_{i}=\left[\sum_{j=1}^{K}\mathbf{t}_{j}g_{c,ji}\right]\textrm{ mod }\Lambda
\]
of the lattice points $\mathbf{t}_{j}$ for $j\in\left[K\right]$. 

Dither vectors $\mathbf{d}_{j}$ are generated independently by a
uniform distribution over the Voronoi region $\mathcal{V}$ of the
coarse lattice $\Lambda$. All dither vectors are available at the
servers. The user transmits 
\[
\mathbf{x}_{j}=\left[\mathbf{t}_{j}-\mathbf{d}_{j}\right]\textrm{ mod }\Lambda.
\]
By \cite[Lemma 7]{ComptForward}, the vector $\mathbf{x}_{j}$ is
uniform over $\mathcal{V}$, so we have the equality $\mathbb{E}[\left\Vert \mathbf{x}_{j}\right\Vert ^{2}]=nP$,
where the expectation is over all dithers. Furthermore, it is argued
in \cite{ComptForward} that there exist fixed dithers that meet the
power constraint $\left\Vert \mathbf{x}_{j}\right\Vert ^{2}\leq nP$. 

The input of Sever $i\in\left[N\right]$ is given by (\ref{eq:reveivedPacket}).
Each server computes 
\[
\mathbf{s}_{i}=\alpha_{i}\mathbf{\tilde{y}}_{i}+\sum_{j=1}^{K}\mathbf{d}_{j}g_{c,ji}.
\]

Let $Q_{f}$ denote the lattice quantizer for the fine lattice $\Lambda_{f}$.
To obtain an estimation of the lattice equation $\mathbf{v}_{i}$,
this vector is quantized onto $\Lambda_{f}$ modulo the coarse lattice
$\Lambda$.

\[
\begin{aligned}\mathbf{\hat{v}}_{i} & =\left[Q_{f}\left(\mathbf{s}_{i}\right)\right]\textrm{ mod }\Lambda\\
 & =\left[Q_{f}\left(\left[\mathbf{s}_{i}\right]\textrm{ mod }\Lambda\right)\right]\textrm{ mod }\Lambda.
\end{aligned}
\]
The following sequence of qualities shows that $\left[\mathbf{s}_{i}\right]\textrm{ mod }\Lambda$
is equivalent to $\mathbf{v}_{i}$ with some added noise terms. 

\begin{equation}
\begin{aligned} & \left[\mathbf{s}_{i}\right]\textrm{ mod }\Lambda\\
 & =\left[\sum_{j=1}^{K}g_{c,ji}\left(\left[\mathbf{t}_{j}-\mathbf{d}_{j}\right]\textrm{ mod }\Lambda+\mathbf{d}_{j}\right)\right.\\
 & \left.\hspace{20bp}+\sum_{j=1}^{K}g_{c,ji}\left(\left(\alpha_{i}-1\right)\mathbf{x}_{j}+\alpha_{i}\mathbf{z}_{j}\right)\right]\textrm{ mod }\Lambda\\
 & =\left[\sum_{j=1}^{K}g_{c,ji}\mathbf{t}_{j}+\sum_{j=1}^{K}g_{c,ji}\left(\left(\alpha_{i}-1\right)\mathbf{x}_{j}+\alpha_{i}\mathbf{z}_{j}\right)\right]\textrm{ mod }\Lambda\\
 & =\left[\mathbf{v}_{i}+\sum_{j=1}^{K}g_{c,ji}\left(\left(\alpha_{i}-1\right)\mathbf{x}_{j}+\alpha_{i}\mathbf{z}_{j}\right)\right]\textrm{ mod }\Lambda.
\end{aligned}
\label{eq:decodefunction}
\end{equation}
By \cite[Lemma 7]{ComptForward}, the pair $\left(\mathbf{v}_{i},\mathbf{\hat{v}}_{i}\right)$
has the same joint distribution as the pair $\left(\mathbf{v}_{i},\mathbf{\tilde{v}}_{i}\right)$,
where $\mathbf{\tilde{v}}_{i}$ is defined as 
\[
\mathbf{\tilde{v}}_{i}\triangleq\left[Q_{f}\left(\mathbf{v}_{i}+\mathbf{z_{eq,}}_{i}\right)\right]\textrm{ mod }\Lambda,
\]
where 
\[
\mathbf{z_{eq,}}_{i}\triangleq\sum_{j=1}^{K}g_{c,ji}\left(\left(\alpha_{i}-1\right)\mathbf{x}_{j}+\alpha_{i}\mathbf{z}_{j}\right),
\]
and $\mathbf{x}_{j}$ is drawn independently and uniformly distributed
over $\mathcal{V}$. By \cite[Lemma 8]{ComptForward}, the density
of $\mathbf{z_{eq,}}_{i}$ can be upper bounded by an i.i.d. zero-mean
Gaussian vector $\mathbf{z}_{i}^{*}$ whose variance $\sigma_{eq,i}^{2}$
approaches 
\[
N_{eq,i}=\left\Vert \mathbf{g}_{c,i}\right\Vert ^{2}N_{0}\left(\alpha_{i}^{2}+\textrm{SNR}\left(\alpha_{i}-1\right)^{2}\right),
\]
as $n\rightarrow\infty$.

The probability of error $\textrm{Pr}\left(\mathbf{\hat{v}}_{i}\neq\mathbf{v}_{i}\right)$
is thus equal to the probability that the equivalent noise leaves
the Voronoi region surrounding the codeword, $\textrm{Pr}\left(\mathbf{z_{eq,}}_{i}\notin\mathcal{V}_{f}\right)$.
Also, we design the fine lattice such that $\Lambda_{f}$ satisfies
AWGN-goodness \cite{poltyrev}, which requires that $\epsilon_{i}=\textrm{Pr}\left(\mathbf{z}_{i}^{*}\notin\mathcal{V}_{f}\right)$
goes to zero exponentially in $n$ as long as the volume-to-noise
ratio is such that

\begin{equation}
\mu\left(\Lambda_{f},\epsilon_{i}\right)\triangleq\frac{\left(\textrm{Vol}\left(\mathcal{V}_{f}\right)\right)^{2/n}}{\sigma_{eq,i}^{2}}>2\pi e.\label{eq:vol-to-noise-ratio}
\end{equation}

Under this condition, $\epsilon_{i}=\textrm{Pr}\left(\mathbf{z_{eq,}}_{i}\notin\mathcal{V}_{f}\right)$
also goes to zero exponentially in $n$. By the union bound, the average
probability of error $\epsilon$ is upper bounded by $\epsilon\leq\sum_{i=1}^{N}\textrm{Pr}\left(\mathbf{z_{eq,}}_{i}\notin\mathcal{V}_{f}\right).$
To ensure that $\epsilon_{i}$ goes to zero for all desired equations,
$\mathcal{V}_{f}$ must satisfy (\ref{eq:vol-to-noise-ratio}) for
all servers with $g_{c,ji}\neq0$. We set $\mathcal{V}_{f}$ such
that the constraint 
\[
\textrm{Vol}\left(\mathcal{V}_{f}\right)>\left(2\pi e\max_{i:g_{c,ji}\neq0}\sigma_{eq,i}^{2}\right)^{n/2}
\]
is always met.

The rate of a nested lattice code is given by $R=\frac{1}{n}\log\frac{\textrm{Vol}\left(\mathcal{V}\right)}{\textrm{Vol}\left(\mathcal{V}_{f}\right)}.$
By (\ref{eq:Def-NSM}), we derive $\textrm{Vol}\left(\mathcal{V}\right)=\left(\frac{P}{G\left(\Lambda\right)}\right)^{n/2}.$
It follows that we can achieve any rates satisfying 
\[
R<\min_{i:g_{c,ji}\neq0}\frac{1}{2}\log^{+}\left(\frac{P}{G\left(\Lambda\right)2\pi e\sigma_{eq,i}^{2}}\right).
\]

Since $\Lambda$ satisfies quantization-goodness \cite{Zamir1996QuantGoodnees}
for $n$ large enough by assumption, we have $G\left(\Lambda\right)2\pi e<\left(1+\delta\right)$
for any $\delta>0$. Knowing that $\sigma_{eq,i}^{2}$ converges to
$N_{eq,i}$, so for $n\rightarrow\infty$, we have $\sigma_{eq,i}^{2}<\left(1+\delta\right)N_{eq,i}$.
Finally, we derive that the rate of the nested lattice code should
be at least 
\[
\begin{aligned}R<\min_{i:g_{c,ji}\neq0}\frac{1}{2}\log^{+}\left(\frac{P}{N_{eq,i}}\right)-\log\left(1+\delta\right).\end{aligned}
\]
Therefore, by choosing $\delta$ small enough, we can approach the
computation rate as close as we desired.

As a result, the servers can make estimates $\mathbf{\hat{v}}_{i}$
of lattice equations $\mathbf{v}_{i}$ with coefficient vectors $\mathbf{g}_{c,1}$,
$\mathbf{g}_{c,2}$,$\ldots$, $\mathbf{g}_{c,N}\in g\left(\mathbb{\mathbb{F}}_{p'}\right)^{K}$
such that $\textrm{Pr}\left(\mathbf{\hat{v}}_{i}\neq\mathbf{v}_{i}\right)<\epsilon$
for $\epsilon>0$ and large $n$ enough as long as 
\[
R<\min_{i:g_{c,ji}\neq0}\frac{1}{2}\log^{+}\left(\frac{P}{\left\Vert \mathbf{g}_{c,i}\right\Vert ^{2}N_{0}\left(\alpha_{i}^{2}+\textrm{SNR}\left(\alpha_{i}-1\right)^{2}\right)}\right)
\]
for some $\alpha_{1},\ldots,\alpha_{N}\in\mathbb{R}$. Finally, using
$\phi^{-1}$ from \cite[Lemma 6]{ComptForward}, each server can produce
estimates of the desired linear combination of messages $\mathbf{\hat{u}}_{i}=\phi^{-1}\left(\mathbf{\hat{v}}_{i}\right)$
such that $\textrm{Pr}\left(\bigcup_{i=1}^{N}\left\{ \mathbf{\hat{u}}_{i}\neq\mathbf{\tilde{u}}_{i}\right\} \right)<\epsilon$
where 
\[
\begin{aligned}\mathbf{\tilde{u}}_{i}=\bigoplus_{j=1}^{K}\tilde{g}_{c,ji}\mathbf{u}_{j}.\end{aligned}
\]

\section*{Appendix B}

Let $f\left(\alpha_{i}\right)$ denote the denominator of the computation
rate (\ref{eq:computation-rate}). Since it is quadratic in $\alpha_{i}$,
it can be uniquely minimized by setting its first derivative to zero.

\[
\begin{aligned}f\left(\alpha_{i}\right) & =\alpha_{i}^{2}+\textrm{SNR}\left(\alpha_{i}-1\right)^{2}\\
\frac{df}{d\alpha_{i}} & =2\alpha_{i}+2\textrm{SNR}\left(\alpha_{i}-1\right)=0\\
\alpha_{MMSE} & =\frac{\textrm{SNR}}{1+\textrm{SNR}}.
\end{aligned}
\]
We plug $\alpha_{MMSE}$ back into $f\left(\alpha_{i}\right)$ and
substituting this into $\log^{+}\left(\frac{P}{\left\Vert \mathbf{g}_{c,i}\right\Vert ^{2}N_{0}f\left(\alpha_{i}\right)}\right)$
yields the desired computation rate.

\end{document}